\documentstyle{article}
\begin{document}
\title{Developing a local deterministic theory to account
for quantum mechanical effects}
\author{Paul P. Budnik Jr.\\
paul@mtnmath.com\\
Mountain Math Software\\
555 Cresci Road\\
Los Gatos, CA 95030}
\date{\today}
\maketitle
\begin{abstract}
Franson showed that Aspect's experiment to test Bell's inequality
did not rule out local realistic theories with delayed determinism.
A class of local, deterministic discrete mathematical models with
delayed determinism is described that may be consistent with existing
experiments.  These are not hidden variables theories in the sense that
they are not theories of particles plus hidden variables. They are
theories of `hidden' distributed information stored holographic like
throughout a space time region. This information cannot be uniquely
associated with individual particles although it determines the results
observed in particle interactions. The classical parameters of an
interaction are determined as {\it focal points} of continuous nonlinear
changes in the wave function and not as discrete events.  In addition
to not violating Bell's inequality this class of theories can in
principle be distinguished from standard quantum mechanics by other
experiments.  These differences and the experimental constraints on
a test of Bell's inequality to discriminate between the existing theory
and this class of models are discussed.
\end{abstract}

\tableofcontents

\section{Locality}

Quantum mechanics is the most accurately verified physical theory
in existence. It provides an extraordinarily
precise description of all manner of physical phenomena.
It is unique among {\it fundamental} physical theories in providing
a statistical description of nature. Many physicists
think the statistical
description is irreducible. This claim,
in contrast to the physical predictions of the theory,
is metaphysical. It cannot be verified experimentally
or proven analytically. There have
been attempts to do so most notably by von Neuman\cite{Neuman}
but none of these
attempts are accepted as correct today. Bell in his refutation of
von Neuman's proof\cite{BellRefute}
suggested that no such general result is possible.
One must impose some additional constraints
such as locality to prove anything like this.
Bell derived an inequality\cite{BellEquality}
that no local realistic theory can violate
{\it subject to certain timing constraints}. Bell showed quantum
mechanics predicts this inequality is violated within the timing
constraints.

There have been many experiments to test Bell's inequality but
only one of these, Aspect's\cite{Aspect},
have attempted to insure the timing
constraints necessary to show a violation of Bell's inequality were
met. This is true even of recently reported
experiments\cite{Rarity1,Brendel1,Brendel2,Kwait1}.
A recent proposal to close all loopholes\cite{Kwait2}
addresses the timing
issues but not adequately as we discuss below.

Franson showed that a local realistic theory
that possessed what he termed delayed determinism
could account for those results\cite{Franson}.
Franson's notion of delayed determinism i.~e. that an event may
not be determined until some time after it has been completed, may
seem strange and unrealistic. However there is no
objective definition of event in quantum mechanics. The unobserved
microscopic events that Franson discusses (such as the emission of
a photon by an excited atom) are hypothetical. It is a mistake
to assume that such events occur as macroscopic events do.
Quantum mechanics only
allows us to compute the probabilities of making observations given
certain initial conditions. What happens between the time we
set up the initial conditions and make an observation is the
{\it terra incognita} of quantum mechanics. We cannot
base the timing in a test of Bell's inequality on the hypothetical
times of hypothetical events.

Franson's objections to Aspect's experiment showed that
there is no {\it objective} criteria in the formalism of the
existing theory for computing the timing in an experimental
test of Bell's inequality. One way to understand this is through
the thought experiment of
Schr\"{o}dinger's cat\cite{CAT}. Schr\"{o}dinger begins his
description of this experiment with: ``One can even set
up quite ridiculous cases.'' A cat
is in a superposition of states such that whether the cat is
alive or dead is not determined (according to the Copenhagen
interpretation) until someone opens the apparatus
and observes the cat. An autopsy of the cat would reveal the time
of death but the time at which it was {\it determined} whether the cat
lives or dies is when the cat is observed which can be much
later than the time of death revealed by the autopsy.
I agree this is a ridiculous example, however
it is consistent with the formalism of
quantum mechanics. There is nothing in that formalism that allows
us to know when macroscopic events are irreversibly determined.
That question is left to interpretations which for the most part
are metaphysical and not subject to experimental tests. Thus there
is no way to decide among them. This problem
applies not only to tests of Bell's inequality but to any experiment
that asks questions about the timing of causal sequences of
{\it macroscopic} events.

If the timing cannot be derived from the formalism of quantum mechanics
or from an interpretation of the theory then
it must be derived from
a competing theory. Developing
such alternatives, even if extremely speculative, is a critical
element in designing tests of Bell's inequality. The timing
constraints I describe in Section~\ref{Sec:timing} apply to
a broad class of alternative theories and not just the class
of models I advocate. These timing constraints are
often assumed by experimenters perhaps without fully realizing that
they cannot be derived from the formalism of the exiting theory.

A recent analysis which claims to describe how to close
all the loopholes in tests of Bell's inequality\cite{Kwait2}
is incomplete in its analysis of the timing
issues. The authors state on page 3210: ``To close this loophole,
the analyzer's settings should be changed {\it after}
the correlated pair has left the source.''
There is no way to know when the pair has left the source
unless one detects them at that point which makes the
experiment impossible. The speed of the process that
generate the photons is only relevant if there is a common trigger
for that process and the changing of the polarizer angles.
Perhaps this is what the authors are suggesting.
The timing can only involve {\it macroscopic} events
such as setting the polarizers or {\it macroscopic effects}
from detecting the photons. The basis for determining the
times of these events must come from a competing theory. The authors
do not discuss this or the need to base timing on purely macroscopic
events.
In Section~\ref{Sec:timing} we describe what must be done to address
the timing issue in practical experiments.

\section{The form of a local realistic deterministic theory}

A local realistic deterministic theory will not violate Bell's inequality
and will provide a deterministic (not statistical) description
of nature. This suggests that it will differ from the existing theory
in specific ways. One would expect a local realistic theory to
exist entirely in physical space as opposed to the Hilbert space
and state space required by the existing theory.
Of course it is reasonable to use any mathematics that works
as a calculating device. However all
physical events occur in physical
space and one should reasonably expect
a {\it fundamental} mechanistic physical
model that accounts for those events
to exist only in physical space.

If probabilities are not irreducible then any violation of
locality must violate special relativity.
The existing predictions escape this only because
the nonlocal effects are `encrypted' with quantum uncertainty.
One cannot tell if an effect goes from A to B or B to A.
The predictions are the same in any relativistic frame of reference.
However any mechanistic process that produces such results can
only be defined in one frame of reference as it must define a unique
direction in which the effect travels. The mathematics
of quantum mechanics is a
mechanistic model and as such must be tied to a particular frame of
reference. In non-relativistic quantum mechanics configuration
space can only be defined
in an absolute frame of reference. In relativistic
quantum mechanics there are
relativistic fields and a nonlocal state model
that is not relativistic. A local realistic theory cannot use
higher dimensional state space to produce such irreducibly nonlocal
effects without being in direct contradiction with special relativity.
Thus is another reason for expecting such a theory to exist entirely
in physical space.

Einstein felt that that quantities that were conserved absolutely
must have an objective existence beyond the probabilities assigned
to them by quantum mechanics\cite{EPR}.
This led
Einstein to think that there was some additional information
(what other have termed hidden variables)
associated with each particle.
This information would then explain, for example,
both the seeming randomness
of observations of a particle's
momentum and the absolute conservation of
momentum.
All attempts at constructing such models
(with the exception of Bohm's explicitly nonlocal
theory\cite{bohm1}) have been unsuccessful.
It seems unlikely that any local model of this type could succeed.
This does not exhaust the universe of models.
The existing theory may represents the average
or statistical behavior of an objectively real physical
wave function. Particles may be secondary effects derivable
from the wave function and its transformations.

I will now describe a class of models that has these characteristics.
I resist calling these hidden variables theories because the
hidden information is distributed throughout space as the
detailed field values at each point. There are not
variables except in the sense that the field value at each point
in the discrete lattice could be considered a variable.

\section{Discretizing the wave equation}

Near the end of his life Einstein
came to suspect that physics cannot
be based on continuous structures. He discussed this in a letter to
Besso quoted on page 467 by Pais\cite{DiscrEinstein}.

\begin{quote}
I consider it quite possible that physics cannot be based on the
field concept, i.~e., on continuous structures. In that case
{\it nothing} remains of my entire castle in the air gravitation
theory included, [and of] the rest of modern physics.
\end{quote}

This insight may be a clue to understanding the nonlinear behavior of
a physical wave function. The simplest model for a local deterministic
physical theory is a field function i.~e. a function defined at each space
time coordinate whose evolution is determined by the previous
field values in
the immediate neighborhood. I think it may be possible to construct
all of physics (including particle theory) from a single simple
discretized finite difference equation. The starting point for
any theory like this must be the classical wave equation
for that equation is universal in physics describing both
electromagnetic effects and the relativistic quantum wave function
(Klein Gordon equation) for the photon.

By `discretized' I mean an equation
that is modified to map integers to
integers. A modification is required
because there is no finite difference
approximation to the wave equation
that can do this. The universality of
the wave function requires that any
discrete model for physics approximates
this continuous model to extraordinary accuracy. Discretizing the
finite difference equation adds a rich combinatorial structure that
has a number of properties that suggest quantum mechanical effects.
Perhaps the most obvious is
that an initial disturbance cannot spread
out or diffuse indefinitely as it does with the continuous equation.
It must break up into independent structures that will continue
to move apart, i.~e., it will eventually become quantized.

We describe how to approximate the wave equation
with a discretized finite difference
equation.
Let $P$ be defined at each point in a 4 dimensional grid.
To simplify the expression for $P_{xyzt}$ we will adopt the following
conventions. Subscripts will be written relative to $P_{xyzt}$ and
will be dropped if they are the same as this point. Thus $P_{t-1}$
is at the same position in the previous time step. $P_{x-1,y-1}$ is
at the same time step and $z$ coordinate and one position less on
both the $x$ and $y$ axes.

The wave equation is approximated by the difference equation:

\begin{eqnarray}
P_{t+1} - 2P + P_{t-1} =
\alpha(P_{x+1} + P_{x-1} + P_{y+1} +
P_{y-1} + P_{z+1} + P_{z-1} - 6P)
\end{eqnarray}

The difference equation discretizes space and
time but not the function
defined on this discrete manifold.
The simplest approach to discretizing
the function values is to constrain
them to be integers. This requires
either that $\alpha$ be an integer or
that some rounding scheme be employed
that forces the product
involving $\alpha$ to be an integer.
The former is not possible since
it does not allow for solutions that
approximate the differential equation.

\section{Properties of the discretized wave equation}

{}From the time symmetry one can conclude that
any solution must either diverge or loop through a repeated
sequence that includes the initial conditions. The restriction to
looping or divergence follows
from the discreteness (there are a finite
number of states) and causality
(each new state is completely determined
by the 2 (or N depending on the differencing scheme) previous
states. The loop must include the initial state because of time
symmetry. At any time one can reverse the sequence of the last 2
(or N) states and the entire history will be repeated in reverse.
Thus any loop must include the initial conditions.

The time required for a given system to repeat an exact sequence of
states based on the number
of possibilities easily makes astronomical
numbers appear minute. However if
there are only a small number of stable
structures and the loops do not need to
be exact but only produce states close
to a stable attractor then we can get a
form of structural conservation
law.

For large field values this model
can approximate the corresponding
differential equation to an arbitrarily
high precision. As the intensity
decreases with an initial perturbation
spreading out in space a limit will be reached
when this is no longer possible.
Thus something like field quantization exists.
Eventually the disturbance will break up into separate structures
that move apart from each other. Each of these structures must
have enough total energy to maintain structural stability. This
may require that they individually
continue to approximate the differential
equation to high accuracy.
Such a process is consistent with quantum mechanics
in predicting field quantization. It differs from quantum
mechanics in
limiting the spatial dispersion
of the wave function of a single photon.
It suggests that the wave function
we use in our calculations models
both this physical wave function and
our ignorance of the exact location
of this physical wave function.

\section{A unified scalar field}

An ambitious goal for this class of models is to
unify all the forces and particles
in nature using a single scalar field
and a simple rule for describing the evolution of that field.
The quantum wave function and the electromagnetic field are identical
in this model as they are in the Klein Gordon equation for a
single photon and the classical electromagnetic field equation.

All energy is electromagnetic. This requires
some way to construct neutral matter from an electromagnetic field.
The Klein Gordon equation for a particle with rest mass presents
an additional problem.
\begin{eqnarray}
\frac{\partial^2\psi}{\partial t^2} =
c^2\nabla^2\psi-\frac{m^2c^2\psi}{\hbar^2}
\end{eqnarray}
This is the classical wave equation with a new term involving
the rest mass of the particle.
How can it be derived
from the same rule of evolution that approximates the classical
wave equation?
This may be possible if there is a high carrier frequency near
the highest frequencies that can exist in the discrete model.
The Schr\"{o}dinger wave equation for particles with rest mass
would represent the
average behavior of the physical wave.
It would be the equation for a wave that modulates the
high frequency carrier.
The carrier itself is not a part of any existing model and
would not have significant electromagnetic interactions with
ordinary matter because of its high frequency.

Such a model may be able to
account for the Klein Gordon equation for a particle
with rest mass.
A high frequency carrier wave
will amplify any truncation effect.
Because of this the differential equation that describes the
carrier envelope is not necessarily the same as the differential
equation that describes the carrier. If the carrier is not
detectable by ordinary means then we will only see effects
from the envelope of the carrier and not the carrier itself.
The minimum time step
for the envelope may involve integrating over
many carrier cycles. If round
off error accumulates during this time in a way that is proportional
to the modulation wave amplitude then we will get an equation in
the form of the Klein Gordon equation.

The particle mass squared factor in the Klein Gordon
equation  can be interpreted as establishing
an amplitude scale.
The discretized wave equation may describe the
full evolution of the carrier and the modulating wave that
is a solution of the Klein Gordon equation. However, since
no effects (except mass and gravity) of the high frequency carrier are
detectable with current technology, we only see the effects
of the modulating wave. No matter how
localized the particle may be it still must have a surrounding
field that falls off in amplitude as $1/r^2$. It is this surrounding
field that embodies the gravitational field.

If discretization is accomplished by truncating the
field values this creates a generalized attractive force. It slows
the rate at which a structure diffuses relative to a solution
of the corresponding differential equation by a marginal amount.
Since the gravitational field is a high frequency electromagnetic
field it will alternately act to attract and repel any bit of
matter which is also an electromagnetic field. Round off error
makes the attraction effect slightly greater and the repulsion slightly
less than it is in solutions of the continuous differential equation.

Because everything is electromagnetic in this model
special relativity falls out directly.
If gravity is a perturbation effect of the electromagnetic
force as described
it will appear to alter the space time metric and an approximation
to general relativity should also be derivable. It is only the
metric and not the space time manifold (lattice of discrete points)
that is affected by gravity.
Thus there is an absolute frame of reference. True singularities
will never occur in this class of models. Instead one will expect
new structures will appear at the point where the
existing theory predicts mass will collapse to
a singularity.

\section{Symmetry in a fully discrete model}

A fully discrete model cannot be completely symmetric as
a continuous model can be. There are ways around this like
using a random lattice but such models implicitly assume a continuous
manifold. In a fully discrete model there must be an absolute
frame of reference and preferred directions in that frame related
to the graininess of the lattice that defines the
space time manifold. One would expect experimental affects
from this absolute frame of reference and perhaps such affects
have already been observed. It is conceivable that the symmetry
breaking that has been observed in weak interactions is a result
of our absolute motion against this manifold and not a break down
of parity.
\section{Dynamically stable structures}

It is likely that the structures an initial disturbance breaks into
will be somewhat
analogous to attractors in chaos theory.
These attractors will be dynamically
stable structures that pass through
similar sequences of states even if
they are slightly perturbed. Such structures will be transformed
to different structures
or `attractors' if they are perturbed sufficiently.
These structures have a form of wave-particle duality. They
are extended fields that transform as structural units.
It is the `structural integrity' of
these `attractors' that may explain
the multi-particle wave function. These structures
can physically overlap. In doing so they loose their individual
identities.
The relationship between the observation
of a particle to earlier observations
of particles in a multi-particle system does not require any
continuity in the existence of these particles. Particles are not
indivisible structures. They are the focal point and mechanism
through which the
wave function interacts and reveals its presence.

It is plausible to expect such a
system will continually be resolving itself into stable structures.
Reversibility and
absolute time symmetry put constraints on what forms of evolution
are possible and what structures
can maintain stability. These may be reflected
in macroscopic laws like the conservation
laws that predict violations
of Bell's inequality. Perhaps we get the correlations
because there is an enormously
complex process of converging to
a stable state consistent with these
structural conservation laws.
It is plausible that at the distances of
the existing experiments
the most probable way this can be
accomplished is through correlations
between observations of the singlet state particles.

In this model {\it isolated} particles are dynamically stable
structures. Multi-particle systems involve the complex dynamics
of a nonlinear wave function that at times and over limited
volumes approximates the behavior of an isolated particle.
Since the existing theory only describes the statistical
behavior of this wave function it is of limited use in gaining
insight into the detailed behavior of this physical wave function.

Consider a particle that emits two photons. In the existing model
there is no event of particle emission. There is a wave function that
gives the probability of detecting either photon at any distance
from the source. Once one of the photons is detected the other
is isolated to a comparatively small region. Prior to detecting either
photon there is a large uncertainty in the position of both photons.
There is even uncertainty as to whether the particle decay occurred and
the photons exist. The existing model gives no idea of what
is actually happening. It only allows us to compute the probability that
we will make certain observations. Some will argue that nothing
is happening except what we observe. In the model I am proposing
there is an objective process involving the emission of two photons.
There is no instant of photon emission. The photons may start
to appear many times and be re-absorbed. At some point the process
will become irreversible and the photons in the
form of two extended wave function structures will move apart.

An observation of either photon localizes both photons in the existing
theory. In my theory there are two localized structures but we do
not know the location of these structures until an observation is made.
For the most part localization effects do not allow discrimination
between my proposal and the standard theory because of
the way the existing theory models the localization of entangled
particles
after an observation.
However in an experiment in which a single particle can diffuse
over an indefinitely large volume there is a difference in the
two theories that is in principle experimentally detectable.
Standard quantum mechanics puts no limit on the distance over which
simultaneous
interference effects from a single particle may be observed.
There will be an absolute fixed limit to this in the class of
theories I am proposing although I cannot quantify what that
limit will be.

Perhaps part of what is so confusing in quantum mechanics
is that it combines classical probability where new
information allows us to `collapse' our model of reality in accord
with an observation and a physical wave function which determines
the probability that there will be a
physical nonlinear
transformation with a focal point at a given location. The existing
theory's failure to discriminate between these two dramatically
different kinds of probability may be one reason why it {\it seems}
to defy conventional notions of causality.

Whether a particular
transformation can complete depends in part on the conservation laws.
Unless there is enough energy to support the new structure and unless
symmetry and other constraints are met a transformation may start to occur
but never complete. One can expect that such incomplete
transformations happen and reverse themselves far more
frequently than do complete transformations. The transformations
that continually start and reverse could be a physical
realization of Feynman diagrams.

A transformation
is a process of {\it converging to stable state} consistent with
the conservation laws. The information that determines the outcome
of this process includes not only the averaged or smoothed wave
function of the existing theory but also the minute details
that result from discretization. This additional hidden information
is not necessarily tied to the particles involved or to their wave
functions in the existing model. It can be anywhere in the light cone
of the transformation process.

\section{The conceptual framework of quantum mechanics}

It has often been suggested that quantum mechanical experiments
produce results that are inconsistent with classical notions of
causality. Bell has proven this is true of the mathematics of
quantum mechanics but the issue is still an open one with regard
to nature. I believe the problem is not with classical ideas of
causality or mathematics but with the conceptual framework with which
we view experimental results. It is important to deal with this issue
explicitly because it is not possible to fully understand the class of models
I propose unless one can think about them in an unconventional
conceptual framework.

Consider
our inability to simultaneously determine a definite position and
momentum for a particle. This result is mathematically related to
our inability to simultaneously fix a position and frequency for
a classical wave. The only wave that has an exact position is an impulse
and that is an integral over all frequencies. We do not think that
this implies any breakdown in classical notions of causality. The behavior
of a classical wave is completely determined just as the behavior of the
quantum mechanical wave function is completely determined.

If point like particles do not exist, it makes no more sense to speak
of their position than it does to speak of the position of a classical
wave. If what we {\it observe} as position is the focal point of a nonlinear
transformation of the wave function then position is a property of
this transformation or interaction and not a property of the particle
itself. If these transformations result from a process of converging to
a stable state consistent with the conservation laws then the information
that determines the detailed characteristics of this transformation may
be spread out over a substantial region of space and may propagate in
ways that are outside of any accepted theory.

Once two particles interact subsequent observations of one
particle puts constraints on observations of the other
even after the particles and their wave functions have
become separated. It is quantum entanglement in the mathematics
of quantum mechanics that is responsible
for violations of Bell's inequality and it is the experimental
phenomenon of quantum entanglement that makes nature
{\it appear} to be inconsistent with classical causality.

The energy and momentum
in a classical wave is distributed
throughout the spatial region occupied
by the wave. If two classical waves
overlap physically there is no clear
way to distribute the energy or
momentum at a
particular point between the two waves.
Once the two wave functions for particles in a multi-particle system
become entangled how do they become disentangled? The wave function
in the existing theory is of limited help if it only represents the
average or statistical behavior of the wave function. If observations
of the particles involve convergence to a stable state consistent
with the conservation laws the the detailed behavior of the physical
wave function is dramatically different from and far more complex
than its average or statistical behavior in the existing model.
Certainly `disentanglement' will occur if the wave functions of
two particles become sufficiently separated. At short distances
tests of Bell's inequality will reveal time delays that allow the
correlations to be determined by information that propagates locally.
At sufficiently great distances the correlations will revert to those
consistent with a local hidden variables model. It will appear as
if the entangled system collapsed spontaneously into two independent
systems. This difference between the existing theory and the class of
models I suggest is not limited to Bell's inequality. Perhaps there
are experimental tests of quantum entanglement that can
more easily
be conducted over large distances to discriminate between these
alternative theories.

\section{Delayed determinism}

Because this model breaks most of the symmetries of the linear
finite difference equation
the classical conservation laws are not enforced
at the local level. There can
be a small discrepancy at any single point
and these discrepancies can
accumulate in a statistically predictable
way. However discreteness and
absolute time symmetry combine to
create a new class of conservation
laws.
The information that enforces
them does not exist at any given point in space or time and cannot
be determined
by a classical space time integral. Instead it is embedded in the
{\it detailed} structure of the state and insures that
the same or similar sequence of states will be repeated.
The local violations of
the conservation laws can never
accumulate in a way that would produce
irreversible events.

Information throughout the light cone of a
transformation puts constraints on what stable
states may
result. A system may start to converge
to two or more stable states but none
of these convergences will complete
unless one of them is consistent
with the conservation laws.
The time of the focal point of this process
(for example the time when
a particle interacts with a detector) and
the time when the event is determined,
i.~e. cannot reverse itself are
not the same thing. Since all interactions
are reversible in this model
the time when an event completes has no
{\it absolute} meaning. It can only
be defined statistically, i.~e.,
the time when the probability that
the event will be reversed is less than some limit.
Quantum mechanics,
because it does not model events objectively,
cannot be used to compute
the probability that an event will be reversed. We must use
classical statistical mechanics. As a practical matter we probably
need to
limit timings to macroscopic measurements where the probability
of the measurement being reversed is negligible.
In the model we propose statistically irreversible
macroscopic events are determined by large number of reversible
microscopic events, i.~e. the nonlinear transformations of
the wave function.
It is important to recognize that use of classical statistical mechanics
to define the occurrence of events implies that quantum mechanics is
an incomplete theory. It is an assumption consistent with the
broad class of theories in which there are {\it objective microscopic
events or processes} that contribute to create macroscopic events.

The distribution of the information that enforces the conservation
laws is not modeled by any accepted theory and is not
limited
by the dispersion of the wave function for
the individual particles.
This information may be distributed throughout
the entire experimental apparatus
including both the particle source and
the detectors. When quantum entanglement
was first discovered there was
some thought that it would disappear once
the wave function for the
entangled particles were spatially
separated\cite{furry,bohm2,broigle,bohm3}.
Aspect's earlier experiments\cite{aspect2}
tested this. These results indicate
that quantum entanglement is not
limited by the spatial dispersion
of the wave function. In a model
like the one we are suggesting the
linear evolution of the wave function
is only part and by far the simplest
part of the picture. Information
that enforces the conservation laws
through quantum entanglement may
evolve in ways that are not remotely
close to linear wave function
evolution. The only
reliable measure of nonlocal quantum entanglement is with
direct {\it macroscopic} measurements of time.

\section{An effective test of Bell's inequality}
\label{Sec:timing}

Bell's inequality is important because
it shows that quantum mechanics
predicts macroscopic violations of
locality. This can only be tested
by suitable {\it macroscopic}
measurements.
To discriminate between the class of theories we are proposing
one must
use statistically irreversible macroscopic events to
measure the timing.
If the probability of reversal is sufficiently low the events
can be treated as if they were absolutely irreversible.
If necessary their probability of
being reversed can be factored into the experimental analysis.
Experimenters often implicitly assume this criteria for the
completion of an event even though it cannot be justified
in the formalism of quantum mechanics.

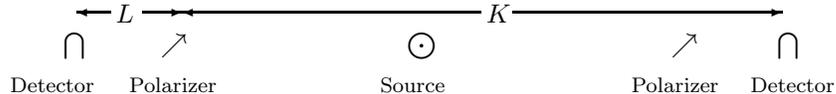
\begin{figure}[h]
\begin{center}
\raggedright
\begin{picture}(400,40)(0,0)

\put(78,30){\vector(-1,0){13}}
\put(80,26){$L$}
\put(90,30){\vector(1,0){15}}

\put(218,30){\vector(-1,0){113}}
\put(220,26){$K$}
\put(230,30){\vector(1,0){102}}

\put(60,15){$\bigcap$}
\put(97,15){$\nearrow$}
\put(190,15){$\bigodot$}
\put(290,15){$\nearrow$}
\put(330,15){$\bigcap$}
\put(40,0){\footnotesize Detector}
\put(85,0){\footnotesize Polarizer}
\put(180,0){\footnotesize Source}
\put(275,0){\footnotesize Polarizer}
\put(320,0){\footnotesize Detector}

\end{picture}
\end{center}
\caption{Typical experiment to test Bell's inequality}
\label{Fig:setup}
\end{figure}

Reported experiments generally involve
a setup such as that shown in Figure~\ref{Fig:setup}.
Quantum mechanics predicts that the correlation between
joint detection will change as a function
of the polarizer (or other experimental apparatus)
settings with a delay given by the time it takes light to
travel the distance $L$.
Most experiments are symmetric. $L$ is the distance from
either polarizer
to the {\it closest} detector.
Locality demands that a change large enough to
violate Bell's inequality can only happen in the time it would
take light to travel the longer distance $K$.
$K$ is the distance from either
polarizer to the {\it more distant} detector.
To show locality is violated one
must show that the delay ($D$) between
when the polarizer settings
are changed and the correlations change
is short enough that $K/D > C$ where $C$ is the speed of light.

It is technically difficult to directly measure
$D$ and none of the
reported experiments do this. Indirect arguments
about $D$ are all questionable.
We have no idea what is happening between the time
the excited state was prepared
and the two detections occurred. Thus we can
make no assumptions about what
is happening microscopically.
This is true both because quantum mechanics
is silent on what is happening
and because these experiments
are testing the correctness of quantum mechanics itself.

To directly measure $D$ requires
that one have a high rate of singlet state
events or a common trigger that controls
these events and the change in polarizer
angles. If this condition is
not met the delay we measure will be dominated
by the uncertainty in when a singlet
state event occurs. After we change
the parameter settings the average
delay we observe will be $D+.5C/r$
where $r$ is the rate of singlet
state events and $D$ is the delay we
want to measure. If $C/r \gg D$ it will
be impossible to accurately
measure $D$. Typical experiments
involve distances of a few meters.
This correspond to expected
values of $D\approx 10$ ns. if locality holds
and $D < 1$ ns. if quantum mechanics is correct. A high
rate of singlet state events or a precise common
trigger for singlet state events and changes
in polarizer angles is necessary to discriminate between
these times.

To show a violation of Bell's inequality one must show the superluminal
transmission of information (at least by Shannon's definition of
information). One must show that a change in polarizer angles
changes the probability of joint detections in less
time than it would take light to travel from either detector to
the more distant analyzer. For this change to be sufficient
to violate Bell's inequality requires that information about
{\it at least one} (we cannot tell which one) polarizer setting
influenced the more distant detector. There must be a macroscopic
record to claim information
has been transferred. It is the time of that record that
must be used in determining if the information
transfer was superluminal.

If one can show superluminal information transfer then one
has a violation of relativistic locality (ignoring the
predeterminism loophole) that is independent of the details of
the experiment. Any attempt to enumerate and eliminate all
loopholes is insufficient because one can never figure out all the
ways that nature might out fox you.

It is worth noting
that the historical roots of these predictions is the assumption
that the wave function changes
{\it instantaneously} when an observation
occurs. This assumption has been built into the mathematics of
quantum mechanics in
a way that creates irreducibly nonlocal operations.
Quantum mechanics insists that there is no
hidden mechanistic process that enforces the conservation laws.
It is this assumption that creates the singlet state entanglement
that enforces conservation laws nonlocally as if by magic
with no underlying mechanism.



\begin{thebibliography}{99}
\bibitem{Neuman} J. von Neuman, {\it The Mathematical Foundations of
Quantum Mechanics} (Princeton University Press, N. J., 1955)
\bibitem{BellRefute} J. S. Bell, Reviews of Modern Physics, {\bf 38},
447 (1966).
\bibitem{BellEquality} J. S. Bell, Physics, {\bf 1},
195 (1964). 
\bibitem{Aspect} A. Aspect, J. Dalibard and G. Roger,
Phys. Rev. Lett., {\bf 49}, 1804 (1982).
\bibitem{Rarity1} J. G. Rarity and P. R. Tapster,
Phys. Rev. Lett., {\bf 64} 2495 (1990).
\bibitem{Brendel1} J. Brendel, E. Mohler and W. Martienssen,
Phys. Rev. Lett., {\bf 66}, 1142 (1991).
\bibitem{Brendel2} J. Brendel, E. Mohler and W. Martienssen,
Europhys. Lett. {\bf 20} 575 (1992).
\bibitem{Kwait1} P. G. Kwait, A. M. Steinberg, and R. Y. Chiao,
Phys. Rev. A {\bf 47} 2472 (1993).
\bibitem{Kwait2} P. G. Kwait, P. H. Eberhard, A. M. Steinberg, and R. Y. Chiao,
Phys. Rev. A {\bf 49} 3209 (1994).
\bibitem{Franson}J. D. Franson, Phys. Rev. D,
{\bf 31}, 2529 (1985).
\bibitem{EPR} A. Einstein, B. Podolsky and N. Rosen, Phys. Rev.,
{\bf 47}, 777 (1935).
\bibitem{CAT} E. Schr\"{o}dinger, Proceedings of the American Philosophical
Society,
{\bf 124}, 777 (1935).
\bibitem{bohm1} D. Bohm, Phys. Rev.,  {\bf 85} 166 (1952).
\bibitem{DiscrEinstein} A. Pais, {\it Subtle is the Lord}, 
    (Oxford University Press, New York, 1982).
\bibitem{furry} W. H. Furry, Phys. Rev. {\bf 49}, 393 (1936).
\bibitem{bohm2} C. Bohm and Y. Aharobov,
Phys. Rev. {\bf 108}, 1070 (1957).
\bibitem{broigle} L. de Broigle, C. R. Acad. Sci.
{\bf 278B}, 721 (1974).
\bibitem{bohm3} D. Bohm and B. J. Hilley,
Nuovo Cimento {\bf B35}, 137 (1976).
\bibitem{aspect2} A. Aspect, P. Grangier and G. Roger,
Phys. Rev. Lett., {\bf 47}, 460 (1981).
\end{thebibliography}
\end{document}